\documentclass[10pt,conference,letterpaper]{IEEEtran}
\IEEEoverridecommandlockouts
\usepackage{cite}
\usepackage{amsmath,amssymb,amsfonts}
\usepackage{algorithmic}
\usepackage{graphicx}

\usepackage{textcomp}
\usepackage{xcolor}
\usepackage[hyphens]{url}
\usepackage[hidelinks]{hyperref}
\usepackage[all]{hypcap}
\usepackage{makecell}
\usepackage{multirow}
\def\BibTeX{{\rm B\kern-.05em{\sc i\kern-.025em b}\kern-.08em
    T\kern-.1667em\lower.7ex\hbox{E}\kern-.125emX}}
\begin{document}

\title{Crucial and Redundant Shares and Compartments in Secret Sharing
\thanks{The authors acknowledge the financial support by the Federal Ministry of Education and Research of Germany in the framework of SoNaTe (project number 16SV7405).}
}

\author{\IEEEauthorblockN{Fabian Schillinger}
	\IEEEauthorblockA{\textit{Computer Networks and Telematics}\\\textit{Department of Computer Science}\\\textit{University of Freiburg}\\
		Freiburg, Germany\\
		\href{https://orcid.org/0000-0001-8771-8290}{\includegraphics[height=1.3\fontcharht\font`\B]{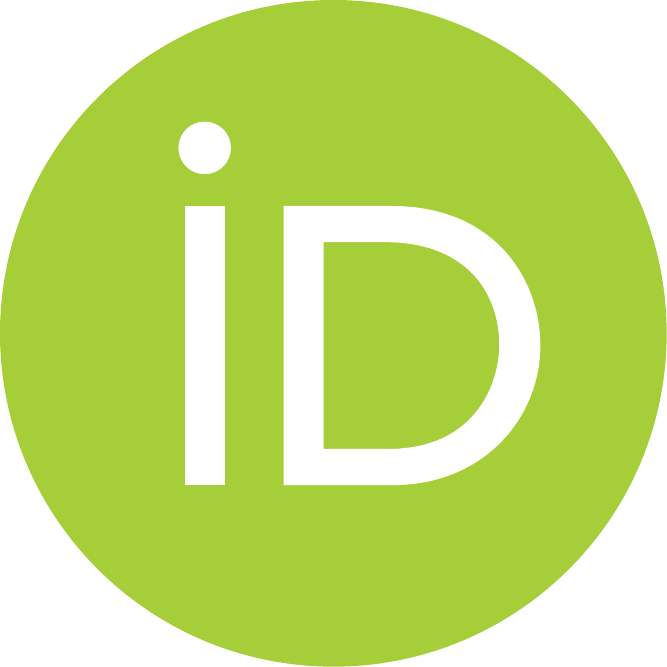} \textcolor{blue}{\url{https://orcid.org/0000-0001-8771-8290}}}
	}\and
	\IEEEauthorblockN{Christian Schindelhauer}
	\IEEEauthorblockA{\textit{Computer Networks and Telematics}\\\textit{Department of Computer Science}\\\textit{University of Freiburg}\\
		Freiburg, Germany\\
		\href{https://orcid.org/0000-0002-8320-8581}{\includegraphics[height=1.3\fontcharht\font`\B]{ORCID} \textcolor{blue}{\url{https://orcid.org/0000-0002-8320-8581}}}}
}
\newtheorem{example}{Example}
\newtheorem{definition}{Definition}
\newtheorem{lemma}{Lemma}
\maketitle

\begin{abstract}
Secret sharing is the well-known problem of splitting a secret into multiple shares, which are distributed to shareholders. When enough or the correct combination of shareholders work together the secret can be restored. We introduce two new types of shares to the secret sharing scheme of Shamir. Crucial shares are always needed for the reconstruction of the secret, whereas mutual redundant shares only help once in reconstructing the secret. Further, we extend the idea of crucial and redundant shares to a compartmented secret sharing scheme. The scheme, which is based on Shamir's, allows distributing the secret to different compartments, that hold shareholders themselves. In each compartment, another secret sharing scheme can be applied. Using the modifications the overall complexity of general access structures realized through compartmented secret sharing schemes can be reduced. This improves the computational complexity. Also, the number of shares can be reduced and some complex access structures can be realized with ideal amount and size of shares.
\end{abstract}


\section{Introduction}
\label{Introduction}
A secret sharing scheme allows a dealer to distribute a secret, like an access code to multiple users, often called shareholders. The parts of the secret, often called shares or shadows can be used to reconstruct or reveal the secret when it is lost or destroyed. A simple secret sharing scheme works as follows: First, the dealer converts the secret into a number $S$ from a Galois field modulo~$p$. Second, it generates a uniformly distributed random number $r_i$ from $GF(p)$ for each but one shareholder and calculates a $r'$, such that the equation ${S=\sum_{i}r_i + r'\mod p}$ holds. Third, it distributes each share $r_i$ or $r'$ to the according shareholder. When all shareholders work together they can calculate the sum of their shares to reveal $S$. This allows distributing a secret in a way, such that no shareholder can calculate $S$ solely from their share. The secret cannot be revealed anymore if any shareholder stops helping. Threshold secret sharing schemes overcome this drawback.
\subsection{Threshold Secret Sharing Schemes}
A $(t,n)$-threshold secret sharing scheme (TSSS) allows a dealer to define some threshold $t$ and to split a secret $S$, into $n$ shares. The shares are distributed to the shareholders. When the threshold is met, i.e. enough shareholders combine their shares, they can reveal $S$. The following scenario can be solved by a TSSS: \begin{example}\label{problem:tsss}To open the vault of a company multiple people have to work together. The company owner ($o$), three managers ($m_1, m_2, m_3$), and three shift leaders ($s_1, s_2, s_3$) each have a private access code for the vault. As soon as two of the seven people enter their code the vault can be opened.\\Here, each of the $n=7$ persons receives a single share, the threshold is $t=2$.\end{example} Secret sharing schemes were first proposed in 1979 by Shamir~\cite{shamir1979share} and Blakley~\cite{blakley1979safeguarding}. In the scheme of Shamir a random polynomial of degree $t-1$ is generated, such that the intersection with the $x$-axis defines the secret. The polynomial is used to calculate $n$ points, which are distributed. When $t$~points are known the secret can be calculated. The scheme is discussed in detail in Section~\ref{Secret Sharing Backgrounds}. In the scheme of Blakley, the secret is a point of intersection of hyperplanes in a $t$~dimensional space. Other approaches, like \cite{cooper1994secret} or \cite{harn2014verifiable} use Latin squares or the Chinese Remainder Theorem.
\subsection{Hierarchical Threshold Secret Sharing Schemes}
Hierarchical threshold secret sharing schemes (HSSS), or multilevel threshold schemes allow organizing shareholders in different groups. Each group is a subgroup of a larger group, where the largest group contains all shareholders. This allows replacing shares of a group by shares out of the parent groups. An HSSS can solve the following scenario: \begin{example}\label{problem:hsss}To open the vault of a company multiple people have to work together. The company owner, three managers, and three shift leaders each have a private access code for the vault. To open the vault three access codes are needed. The owner and managers have higher positions, so at least one of the access codes must be provided by them. The vault can be opened for example if two of the managers, together with a shift leader enter their codes, or if the owner, one manager, and one of the shift leaders enter their codes.\\
	One group consists of all the shift leaders. It is the only subgroup of the group containing all persons. \end{example}
Multiple approaches for HSSS are presented in \cite{simmons1988really,shima2017hierarchical,tassa2004hierarchical,farras2012ideal,harn2014multilevel,tentu2013ideal,ballico2007hierarchical,zhang2008sure}. 
\subsection{Compartmented Secret Sharing Schemes}
In compartmented secret sharing schemes (CTSS) the shareholders are grouped in different compartments. Each compartment receives a share through a secret sharing scheme. Each share is used as a new secret and distributed, using another secret sharing scheme to the users in the compartment. This allows to retrieve a secret in a conjunctive CTSS if in every compartment the secret is retrieved, and then the secrets are combined. In a disjunctive CTSS, only in a specific number of compartments, the shares have to be revealed to calculate the secret. A CTSS can solve the following scenario: \begin{example}\label{problem:ctss}To open the vault of a company multiple people have to work together. The company owner, three managers, and three shift leaders each have a private access code for the vault. The company owner and the three managers form the higher management, whereas the shift leaders form the lower management. \\In a conjunctive CTSS two people from the higher management and two people from the lower management have to enter their codes to open the vault. \\In a disjunctive CTSS, either two persons from the higher management, or two persons from the lower management suffice to open the vault.\\There are two compartments: $C_{h}$ containing the persons from the higher management, with threshold $t_{h}=2$. The other one $C_{l}$, with threshold $t_{l}=2$ contains the persons from the lower management. The key for the vault is distributed two both compartments, using a TSSS with threshold $t_{c}=2$ in the conjunctive case and $_{d}=1$ in the disjunctive case. Both compartments distribute their share to the shareholders using another TSSS.\end{example}
It is possible to construct far more complex access structures, by using a CTSS. In \cite{benaloh1990generalized} it is shown, that every CTSS may be used for general access structures. Shareholders might receive more than one share in those schemes. The approach discussed in \cite{lin2009ideal_a} can be used to generate both, HSSS and CTSS. Other approaches are presented in \cite{ghodosi1998secret,iftene2005compartmented,simmons1988really}.

\subsection{Weighted Threshold Secret Sharing Schemes}
In a weighted threshold secret sharing schemes (WTSS) each shareholder has a specific weight. If the sum of weights of the combined shares is larger than the threshold value, the secret can be revealed. Single shares of shareholders with a higher weight, therefore, can replace multiple shares of shareholders with lower weights. A WTSS can solve the following scenario: \begin{example}\label{problem:wtss}To open the vault of a company multiple people have to work together. The company owner, three managers, and three shift leaders each have a private access code. The company owner can open the vault on his own, at least two of the managers can open the vault together, or two shift leaders together with an additional manager can open the vault.\\
	The threshold for this scheme is $t=5$. The access code of the company owner has a weight of 5, the codes of the managers each have a weight of 3, the codes of the shift leaders each have a weight of 1. The combinations of different weighted access codes results in the described access structure.\end{example}
Some WTSS are presented in \cite{morillo1999weighted,beimel2005characterizing,beimel2006monotone,padro2000secret,iftene2005weighted,belenkiy2008disjunctive,Tassa2009}.
\subsection{General Access Secret Sharing Schemes}
The previous schemes were able to map specific access structures efficiently. In \cite{ito1989secret}, the multiple assignment scheme is proposed, where multiple shares are assigned to each shareholder. With this approach, general access structures can be realized, with the downside, that in the worst case a total of ${\mathcal{O}(n\cdot2^n)}$ shares have to be distributed to $n$ shareholders. A general access secret sharing scheme allows mapping every possible subgroup of shareholders into the access structure. 
Multiple schemes improve the efficiency by reducing the number of shares or the size of each share: In~\cite{benaloh1990generalized} a method is presented, which allows to construct general access structures from nested combinations of conjunctions, disjunctions, and threshold-functions. This results in multiple shares per shareholder when a nesting is used. Linear Block Codes are used in~\cite{bertilsson1992construction}. To create the shares a generator matrix of size $~ U \times U$ for $U$ shareholders is used. Each shareholder receives a share. In~\cite{dawson1994breadth} each shareholder receives a share for each minimal authorized subset he is a part of. A authorized subset is a set of shareholders, which can compute the secret. It is minimal if no subset is sufficient to compute the secret. The amount of shares can be reduced in some cases. Then the same polynomial to calculate the shares is used in multiple subgroups. The approach is extended in~\cite{tan1999general}, such that each shareholder $s$, which received $t_s$ shares in~\cite{dawson1994breadth} receives an interpolating polynomial of degree $(t_s-1)$. 
\subsection{Our Contribution}
\label{Introduction Our Contribution}
Various scenarios can be mapped by using the displayed methods of secret sharing. Still, there are some scenarios where either, the amount of distributed shares, the computational complexity or the complexity of the secret sharing scheme can be reduced, by introducing two special kinds of shares: Crucial shares are always needed to reveal a secret, regardless of whether the threshold value is achieved, or not. Redundant shares are pairs or sets of shares, where the first redundant share helps in revealing the secret, but any additional redundant share does not help any further. 
Our contribution consists of modifications to the secret sharing scheme of Shamir, and the compartmented secret sharing scheme proposed by Simmons~\cite{simmons1988really}, which is based on Shamir's scheme. The modifications allow constructing general access schemes while many of the good properties of Shamir's scheme, like, easy implementation and high understandability can be retained. In some cases even ideally can be achieved by using our modifications.
\subsection{Exemplary Use-Cases for our Contribution}
\label{Our Contribution Use-cases}
The following examples show use-cases where our modifications to the TSSS of Shamir and the CTSS of Simmons can help in creating schemes which are more efficient than the state-of-the-art schemes. \begin{example}\label{problem:normal secret sharing} To open the vault of a company multiple people have to work together. The company owner ($o$), the head of security ($\mathit{sec}$), three managers ($m_1, m_2, m_3$), and three shift leaders ($s_1,s_2,s_3$) each have a private access code for the vault. Any combination of four codes can open the vault, but in any case, the codes of the owner and the head of security have to be two of them, and the remaining two codes cannot be both from shift leaders.
\end{example} Example~\ref{problem:normal secret sharing} can be mapped with our modified secret sharing scheme as follows: the managers $m_1$, $m_2$, and $m_3$ each receive a normal share from the dealer, the shift leaders $s_1$, $s_2$, and $s_3$ each receive a mutual redundant share, and both the owner $o$ and the head of security $\mathit{sec}$ receive crucial shares. The threshold is $t=2$, the number of different shares is $n=4$. By using a state-of-the-art scheme, like the one presented in \cite{dawson1994breadth}, more than 4 shares have to be distributed: the disjunctive form describing the various group composition that allow to open the vault contains 12 different clauses like $(o \wedge \mathit{sec} \wedge m_1 \wedge m_2)$, each containing 4 literals. For every clause a different \mbox{($4,4$)-TSSS} is used to determine the access. Therefore, in total 48 shares have to be distributed. This can be reduced to 26 shares because $o$ and $\mathit{sec}$ are present in every clause. Therefore, $o$ and $\mathit{sec}$ receive shares for a \mbox{$(2,2)$-scheme} with secret $S_1$. The 12 remaining clauses $(m_1 \wedge m_2)$, $(m_1 \wedge m_3)$, $(m_2 \wedge m_3)$, $\dots$ are mapped by \mbox{$(2,2)$-TSSS}, each having the same secret $S_2$. Access to the vault then is calculated through $S_1+S_2$.


%
%
%
%
%
%

\begin{example}\label{problem:compartmented secret sharing} To open the vault of a company multiple people have to work together.There are three departments, each with a manager and a deputy, and a group of staff.
The vault can be opened if the majority of staff in two departments are supported by two managers or deputies from two departments.
\end{example}
Example~\ref{problem:compartmented secret sharing} can be mapped with our modified CTSS as follows: The three managers and their deputies form a crucial compartment, where the secret is shared in a $(2,3)$-secret sharing. Each manager receives a normal share and the respective deputy each receives a redundant share. Therefore, the secret in the crucial compartment can be calculated if two managers, one manager, and another department's deputy or two deputies combine their shares. The staff of each department~$i$ forms a compartment $C_i$, where each secret is the share of a $(2,3)$-secret sharing. The secret of a compartment $C_i$ is distributed through a $(\lfloor \frac{ |C_i|}{2}\rfloor+1,|C_i|)$-secret sharing. This allows computing a compartments secret if a majority of staff provides their shares. If two of the compartments provide their shares, together with the crucial share the secret can be computed. Here, in total 6+$|C_1|+|C_2|+|C_3|$ shares, one for every person, have to be distributed. By using state-of-the-art schemes, again a significantly higher number of shares have to be distributed: in a department with $|C_i|$ persons and a threshold of $t_i = \lfloor \frac{ |C_i|}{2}\rfloor+1$, there can  occur $\binom{|C_i|}{t_i}$~different combinations of persons. In Example~\ref{problem:compartmented secret sharing} two out of three such departments have to be calculated. In addition, there are 12 different combinations of managers and deputies, which further increase the number of secret sharing schemes needed to cover all possible combinations of persons that can open the vault. Table~\ref{table_example} gives a summary of all examples.
\begin{table*}[!t]
	\caption{The table is a brief summary of the examples from Section~\ref{Introduction}. The abbreviations are as follows: company owner~($o$), managers~($m_i$), deputies~($d_i$), shift leaders ($s_i$), head of security ($\mathit{sec}$), department ($\mathit{dept}_i$), staff of a department ($S_i$).}
	\label{table_example}
	\centering
\begin{tabular}{|c|c|c|c|c|}
	\hline
	\textbf{Example \#} & \textbf{Type} & \textbf{Shareholders} & \textbf{Description of Allowed Combination} &  \textbf{Example Combination} \\
	\hline
	&&&&\\[-1.2ex]
	\hline
	&&&&\\[-1.1ex]
	Example~\ref{problem:tsss} & TSSS & $\{o,m_1,m_2,m_3,s_1,s_2,s_3\}$ & any two shareholders &  $\{o,m_1\},\dots$ \\[1.1ex]
	\hline
	Example~\ref{problem:hsss} & HSSS & \makecell{$G_1=\{o,m_1,m_2,m_3\}$, \\ $G_2=\{s_1,s_2,s_3\}$} & \makecell{any three shareholders from $G_1 \cup G_2$, \\at least one must be from $G_1$} & $\{o,m_1,m_2\},\dots$ \\
	\hline
	
	\multirow{2}{*}{Example~\ref{problem:ctss}} & \multirow{2}{*}{CTSS} & \multirow{2}{*}{\makecell{$G_1=\{o,m_1,m_2,m_3\}$,\\$G_2=\{s_1,s_2,s_3\}$}} & conjunctive CTSS: two from $G_1$ \textsc{AND} two from $G_2$ & $\{o,m_2,s_1,s_3\},\dots$ \\ \cline{4-5} &  & & disjunctive CTSS: two from $G_1$ \textsc{OR} two from $G_2$& $\{o,m_1\},\{s_2,s_3\},\dots$ \\ 
	\hline
	Example~\ref{problem:wtss} & WTSS & \makecell{$G_1=\{o\},G_2=\{m_1,m_2,m_3\}$,\\ $G_3=\{s_1,s_2,s_3\}$} & \makecell{one from $G_1$ \textsc{OR} two from $G_2$ \textsc{OR}\\ (two from $G_3$ \textsc{AND} one from $G_2$)} & $\{o\},\{m_1,s_1,s_3\},\dots$ \\
	\hline
	Example~\ref{problem:normal secret sharing} & \makecell{modified \\ TSSS} & \makecell{$G_1=\{o,\mathit{sec}\},G_2=\{m_1,m_2,m_3\}$,\\ $G_3=\{s_1,s_2,s_3\}$} & \makecell{all from $G_1$ \textsc{AND} \\(($m_A$ \textsc{AND} $m_B$) \textsc{OR} ($m_A$ \textsc{AND} $s_B$))\\ with $A\neq B$}& \makecell{ $\{o,\mathit{sec},m_1,m_2\}$,\phantom{$\dots$}\\$\{o, \mathit{sec}, m_2,s_1\},\dots$} \\
	\hline
	Example~\ref{problem:compartmented secret sharing} & \makecell{modified \\CTSS} & \makecell{$\mathit{dept}_i=\{m_i, d_i, S_i=\{\mathit{st}_{i,j}\}\}$\\ with $i\in \{1,2,3\}$} & \makecell{majority from $S_A$ \textsc{AND} majority from $S_B$ \textsc{AND} \\($m_C$ \textsc{OR} $d_C$) \textsc{AND} ($m_D$ \textsc{OR} $d_D$)\\ with $A\neq B$, $C\neq D$} & \makecell{$\{S'_1, S'_2$, $m_1$, $d_3\}$\\ with $S'_1 \subseteq S_1, S'_2 \subseteq S_2$\\ and $|S'_1|> 0.5|S_1|$,\\ $|S'_2|> 0.5|S_2|$}\\
	\hline
\end{tabular}
\end{table*}

\subsection*{Organization of the Paper}
The paper is structured as follows: Section~\ref{Secret Sharing Backgrounds} describes the backgrounds of secret sharing schemes and explains the used schemes. In Section~\ref{Our Contribution} our contributions are described, Section~\ref{Our Contribution Secret Sharing} states the modifications to the threshold secret sharing scheme of Shamir, whereas in Section~\ref{Compartmented Secret Sharing} the modifications to compartmented secret sharing schemes are discussed. Further, the implications for realizing general access structures are shown. Finally, Section~\ref{Conclusions} concludes the work and gives an outlook on future work.



\section{Secret Sharing Backgrounds}
\label{Secret Sharing Backgrounds}
\begin{definition}[$(t,n)$-Secret Sharing Scheme]
	 In a $(t,n)$-secret sharing scheme $n$~shares ${\mathcal{S}=\{S_1,\dots,S_n\}}$ of a secret $S$ are distributed to a set ${\mathcal{U}=\{U_1,\dots,U_n\}}$ of users. Then, any set $\mathcal{A}$ of shares, with $|\mathcal{A}|\geq t$ can compute $S$.
\end{definition}
\begin{definition}[Access Structure]
	 An access structure is the family $\mathcal{A'}$, with ${\{\mathcal{A'} \subseteq \mathcal{A}: \mathcal{A'}~\text{can reconstruct}~S\}}$ for a secret sharing scheme with secret $S$ and shares $\mathcal{A}$.
\end{definition}
The schemes of Shamir and Blakley are called threshold secret sharing schemes because any subset of shares that reaches the threshold value can reveal the secret. A general access secret sharing scheme allows constructing all access structures.
\begin{definition}[Perfect Secret Sharing Scheme]
	A secret sharing scheme is called perfect, if no set $\mathcal{A'}$ of shares allows learning anything about the secret, if $\mathcal{A'}$ is not in the access structure of the secret sharing scheme.
\end{definition}
This means that correctly guessing the secret $S$ with less than $t$ shares in a $(t,n)$-secret sharing scheme or any set of shares which is not part of the access structure has the same probability as guessing $S$ without a single share. 
\begin{definition}[Ideal Secret Sharing Scheme]
	A secret sharing scheme is called ideal if it is perfect and the size of each shareholders share is in the same domain as the secret. 
\end{definition}
Any scheme where a shareholder receives more than one share cannot be ideal. In the following, we modify the secret sharing scheme of Shamir. This scheme is well-known, easy to understand and to implement. The idea is to generate a polynomial ${f(x)=S+a_1\cdot x+a_2\cdot x^2+\cdots+a_{t-1}\cdot x^{t-1} \mod p}$. $S$ is the secret and the coefficients $a_1,a_2,\dots a_{t-1}$ are uniformly distributed random variables, each from $GF(p)$, the Galois field of order $p$. Each shareholder then receives a point $S_i= f(x_i) \mod p$. 
$S$ can be computed by solving (\ref{eq:lineares gleichungssystem shamir}), with at least $t$ linearly independent combinations of $x_i$ and $S_i$:
\begin{equation}
	\label{eq:lineares gleichungssystem shamir}
	\resizebox{\hsize}{!}{$\displaystyle
\begin{array}{ccc} 
		S+a_1\cdot x_1+a_2\cdot x_1^2+\cdots+a_{t-1}\cdot x_1^{t-1} & = & S_1 \pmod{p} \\ 
		S+a_1\cdot x_2+a_2\cdot x_2^2+\cdots+a_{t-1}\cdot x_2^{t-1} & = & S_2 \pmod{p}\\ 
		\vdots & \vdots & \vdots\\
		S+a_1\cdot x_n+a_2\cdot x_n^2+\cdots+a_{t-1}\cdot x_n^{t-1} & = & S_n \pmod{p}\\ 
	\end{array}$
}
\end{equation}
Any set, containing $t$ shareholders can obtain $S$. The secret sharing scheme of Shamir is ideal \cite{brickell1989some}. For reconstruction of the secret, (\ref{eq:lineares gleichungssystem shamir}) can be used, but there are faster reconstruction procedures. Using the Lagrange polynomial interpolation~\cite{waring1779vii} the secret can be computed if the shareholders $U_i$ calculate the following sum:
\begin{equation}
	\label{eq:lagrange interpolation}
	S = \sum_i \left(S_i\cdot \prod_{i\neq j} \frac{-x_j}{x_i-x_j} \mod p\right)
\end{equation}
\begin{lemma}
	\label{lemma1}
	Calculating shares in Shamir's scheme is in ${\mathcal{O}(t\cdot n)}$.
\end{lemma}
\begin{IEEEproof}
	For generating shares $t-1$ random variables have to be chosen and the polynomial $f(x)$, of degree $t-1$, has to be evaluated. Using Horner's method~\cite{horner1833new}, $t-1$~multiplications are needed for a single share. Therefore, for computing $n$~shares, in total, $t-1+n\cdot (t-1)< t+t\cdot n$ calculations are needed.
\end{IEEEproof}
\begin{lemma}
	\label{lemma2}
	Revealing $S$ in Shamir's secret sharing scheme is in $\mathcal{O}(t^2)$.
\end{lemma}
\begin{IEEEproof}
	A shareholder $i$ has to calculate the product of the share $S_i$ and $(t-1)$ times the given fraction. Therefore, $t$~shareholders have to calculate $t\cdot (t-1)< t^2$ products.
\end{IEEEproof}
\begin{definition}[Compartmented Threshold Secret Sharing Scheme]
	In a compartmented threshold secret sharing scheme with threshold $t$ the users $\mathcal{U}$ are partitioned into compartments $\mathcal{C}=\{C_1,\dots,C_m\}$, such that $\mathcal{U}=\bigcup_{i=1}^{m}C_i$. The Secret $S$ can be computed, if a set $\mathcal{A}$ of compartments, with $|\mathcal{A}|\geq t$ combine their shares. Each compartment $C_i$, can distribute the share $S_i$ in a $(t_i,|C_i|)$-secret sharing scheme to the $|C_i|$ users.
\end{definition}
In the following, the CTSS based on Shamir's scheme, as proposed by Simmons is used. In the scheme a secret $S$ is divided into shares $S_i$ by generating a polynomial $f(x)$ and then calculating a point $f(x_i)$ for each compartment $C_i$. In each compartment, then, another $(t,n)$-secret sharing scheme is applied. I.e., each share $S_i$ is a new secret inside the compartment, which is distributed to the users hold by the compartment. 
To calculate the secret, in $t$ compartments the users have to combine their shares using the Lagrange polynomial interpolation shown in (\ref{eq:lagrange interpolation}). With the $t$~shares another polynomial interpolation has to be calculated to find the secret $S$.
\section{Our Approach}
\label{Our Contribution}
In the following, some modifications to Shamir's scheme are introduced. Later, the respective modifications to the CTSS proposed by Simmons are discussed.
\subsection{Secret Sharing}
\label{Our Contribution Secret Sharing}
In Shamir's secret sharing scheme each share has the same impact. Therefore, in a $(2,4)$-secret sharing scheme with the shares $S_1$, $S_2$, $S_3$, and $S_4$ all sets $\{S_i,S_j\}$, with $i\neq j$ can retrieve the secret. The following modification allows to restrict the access group:
\begin{definition}[Crucial Share]
	\label{definition:crucialshare}
	A share $R$ is called crucial, if there exists no set $\mathcal{A}'$ of shares, with $R\notin \mathcal{A}'$, such that $S$ can be computed. 
\end{definition}
By defining $S_1$ as a crucial share the access group can be restricted to the following: $\{\{S_1,S_2\}, \{S_1,S_3\}, \{S_1,S_4\}\}$.
\begin{lemma}
	\label{lemma crucial share}
	Any number $r$ of crucial shares can be introduced to the secret sharing scheme of Shamir, when $S$ in the polynomial is replaced by some $S'$, with ${S' = S+ \sum_{i=1}^{r}R_i\mod p}$, and all $R_i$ are drawn uniformly random from $GF(p)$. 
\end{lemma}
\begin{IEEEproof}
	Finding the value $S'$ of the modified polynomial is the same as in Shamir's scheme. It can be found by using $t$ linearly independent combinations of $x_i$ and $S_i$ to solve the following equation system:
	\begin{equation}
		\resizebox{\hsize}{!}{$\displaystyle
	 \begin{array}{ccc} 
	S'+a_1\cdot x_1+a_2\cdot x_1^2+\cdots+a_{t-1}\cdot x_1^{t-1} & = & S_1 \pmod{p} \\ 
	S'+a_1\cdot x_2+a_2\cdot x_2^2+\cdots+a_{t-1}\cdot x_2^{t-1} & = & S_2 \pmod{p}\\ 
	\vdots & \vdots & \vdots\\
	S'+a_1\cdot x_n+a_2\cdot x_n^2+\cdots+a_{t-1}\cdot x_n^{t-1} & = & S_n \pmod{p} \\ 
	\end{array}  $
}
\end{equation}
	Less than $t$ linearly independent combinations results in an underconstrained equation system. More than $t$ combinations do not help in finding $S$, because every $R_i$ is independent. When $S'$ is found $S$ can be computed by subtracting all crucial shares: ${S = S'- \sum_{i=1}^{r}R_i \mod p}$. 
	If a crucial share $C$ is missing only a ${S'' = S+C\mod p}$, with $C$ from $GF(p)$ can be computed. Then, $S$ can be any number from $GF(p)$. Therefore, all crucial shares are needed to calculate $S$.
\end{IEEEproof}
\begin{lemma}
	The modified secret sharing scheme with crucial shares remains perfect.
\end{lemma}
\begin{IEEEproof}
	Calculating the secret $S$ in the modified scheme consists of two problems: The first problem is to find $S'$ for ${f'(x) = S'+a_1x+a_2x^2+\cdots+a_{t-1}x^{t-1}\mod p}$. Every set of combinations of $x_i$ and $S_i$, which is smaller than $t$, leads to infinitely many possible polynomials. The other problem is finding $S$ from $S'$, which is moving the correct polynomial vertically. Having more than $t$ linearly independent combinations of $x_i$ and $S_i$ does not help in computing $S$, because the crucial shares $R_i$ are independent of the points of the polynomial. With less than $r$ crucial shares some ${S'' = S+C'\mod p}$, where $C'$ can be any number from $GF(p)$ is left.  Finding $S$ by using $S''$ in a meaningful way, therefore is not possible. Guessing $S$ from this point is as effective, as guessing it without any knowledge.
\end{IEEEproof}
\begin{lemma}
	The modified secret sharing scheme with crucial shares remains ideal.
\end{lemma}
\begin{IEEEproof}
	The modified scheme is perfect. Further, each value $R_i$ is a uniformly chosen random number out of $GF(p)$. The resulting points, when evaluating the polynomial remain in $GF(p)$. Therefore all shares and the secret are from the same domain. 
\end{IEEEproof}
Another way of restricting the access group is by making shares less important than others. In a $(2,4)$-scheme with shares $S_1$, $S_2$, $S_3$, and $S_4$ the access group can be reduced to $\{\{S_1,S_3\}$, $\{S_1,S_4\},\{S_2,S_3\},\{S_2,S_4\}\}$ by not allowing to appear $S_1$ and $S_2$, or $S_3$ and $S_4$ in the same set. This can be achieved by redundant shares:
\begin{definition}[Redundant Share]
	\label{definition:redundantshare}
	Two shares $R$ and $Q$, in a $(t,n)$-secret sharing scheme, are called redundant, if there exists no set $\mathcal{A'}$ of shares, with $R, Q \in \mathcal{A'}$ and $|\mathcal{A'}| = t$, which can compute $S$.
\end{definition}
The definitions for crucial shares and redundant shares can be modified to apply to shareholders instead of shares. But, as shareholders are allowed to hold multiple shares, where the types of shares can be different, the definition may be equivocal. For example, if a shareholder holds a crucial and a redundant share, it can be confusing to call the shareholder crucial or redundant. 
\begin{lemma}
	\label{lemma introducing redundant shares}
	Redundant shares can be introduced to the scheme of Shamir by distributing the same share multiple times to different shareholders.
\end{lemma}
\begin{IEEEproof}
	For finding $S$ in Shamir's scheme least $t$ linearly independent combinations of $x_i$ and $S_i$ are needed. Any set of $s=t$ shares with at least two shares $j,k$ with $x_j=x_k$, $S_j=S_k$, and $j\neq k$ has at most $t-1$ linearly independent shares. 
\end{IEEEproof}
This allows introducing two or more redundant shares having the same value. Multiple redundant shares with different values can be used in the modified secret sharing scheme. We call any number of redundant shares corresponding to the same value mutual redundant shares.
\begin{lemma}
	The modified secret sharing scheme with redundant shares remains perfect.
\end{lemma}
\begin{IEEEproof}
	Redundant shares are copies of normal shares. Knowing multiple mutual redundant shares results in the same amount of linearly independent pairs $x_i$, $S_i$ as having a single redundant share.
\end{IEEEproof}
\begin{lemma}
	The modified secret sharing scheme with redundant shares remains ideal.
\end{lemma}
\begin{IEEEproof}
	Redundant shares are copies of normal shares, therefore they are out of $GF(p)$, as well as the secret $S$.
\end{IEEEproof}

The generation of shares in the modified secret sharing scheme works comparable to Shamir's scheme. At first, all crucial shares $R_i$ and coefficients $a_i$ are chosen uniformly random from $GF(p)$. Then, using these values the polynomial ${f'(x) = S'+a_1x+a_2x^2+\cdots+a_{(t-r-1)}x^{(t-r-1)} \mod p}$, with ${S' = (S +  \sum_{i=1}^{r}R_i )\mod p}$ is generated.\\For the distribution of shares, each shareholder $i$ of a normal share receives a value corresponding to the evaluation ${f'(x_i) \mod p}$ of the polynomial. All shareholders $i$ of crucial shares receive the according value $R_i$. Finally, shareholders of redundant shares receive a copy of the corresponding evaluation of $f'(x)$. \\Then, the secret $S$ can be found by solving the equation ${S = \sum_i \left(S_i\cdot \prod_{i\neq j} \frac{-x_j}{x_i-x_j} \mod p\right) - \sum_{k=1}^{r} R_k \mod p}$.  $x_j$, $x_i$, and $R_k$ are known to the shareholders.
\begin{lemma}
	\label{lemma runtime sharing}
	Calculating shares in the modified scheme is in $\mathcal{O}(t\cdot n)$.
\end{lemma}
\begin{IEEEproof}
	$r$ crucial shares and $t-r-1$ coefficients $a_i$ have to be drawn  because each crucial share reduces the degree of the polynomial $f'(x)$ by one. A normal share $S_i$ is calculated by evaluating  $f'(x)$, which needs $(t-r-1)$ multiplications according to Horner's method. Therefore, calculating ${(n-r-d)}$ normal shares and $r$ crucial shares needs $r+(t-r-1)+(n-r-d)\cdot (t-r-1) < t+t\cdot n$ calculations. For any number of mutual redundant shares only one needs $f'(x)$ to be calculated. All others are copied and do not need calculations.
\end{IEEEproof}
\begin{lemma}
	\label{lemma runtime reconstruction}
	Reconstruction of $S$ in the modified scheme is in $\mathcal{O}(t^2)$.
\end{lemma}
\begin{IEEEproof}
	Each shareholder $U_i$ contributing a normal share or a unique redundant share has to calculate the product of the share $S_i$ and $(t-r-1)$ times the given fraction. Afterwards, the sum of crucial shares $\sum_{k=1}^{r} R_r \mod p$ is subtracted. Therefore, the $(t-r)$ shareholders with a normal share have to calculate in total $(t-r)\cdot (t-r-1) < t^2$ products, this equation holds because $t \geq r \geq 0$.
\end{IEEEproof}
The modified scheme allows using multiple crucial and redundant shares at the same time. Therefore, the access structure can be more flexible. Crucial shares in some way are a contradiction to the initial ideas behind secret sharing. Secret sharing can be used to retrieve a secret which was lost or forgotten. The possibility to lose the secret increases again, when crucial shares are used, because they can be lost, or shareholders might become malicious and stop helping in revealing the secret.
\subsection{Compartmented Secret Sharing}
\label{Compartmented Secret Sharing}
With modifications to compartmented threshold secret sharing schemes the downside of using crucial shares can be offset, because similar to the previous chapter crucial compartments can be introduced, where a compartment is on the one hand needed for the reconstruction of the secret, but on the other hand multiple users help in reconstructing the share held by the compartment. Therefore, the following definitions are similar to the ones of secret sharing.
\begin{definition}[Crucial Compartment]
	A compartment $R$ is called crucial, if there exists no set $\mathcal{A}'$ of shares of compartments, with $R\notin \mathcal{A}'$, such that $S$ can be computed. 
\end{definition}
Crucial compartments can be introduced to the scheme similar to the method shown in Lemma~\ref{lemma crucial share}, by replacing $S$ with a $S' = S+\sum_{i=1}^{r} R_i \mod p$, where $R_i$ are the secrets of the crucial compartments. The secret sharing scheme inside a crucial compartment is independent of the one used in the outer scheme. Therefore, it can be another CTSS or a modified Shamir scheme.
\begin{definition}[Redundant Compartment]
	Two compartments $R$ and $Q$ are called redundant, if there exists no set $\mathcal{A'}$ of shares of compartments, with ${R, Q \in \mathcal{A'}}$ and $|\mathcal{A'}| = t$, which can compute $S$.
\end{definition}
Additionally, redundant compartments can be introduced. Similar to Lemma~\ref{lemma introducing redundant shares}, every redundant compartment receives the same secret, which leads to the case that only one of the mutual redundant compartments can help in revealing the secret. Again, the scheme inside the compartment is independent. Therefore, every redundant compartment can use another secret sharing scheme, a different number of shareholders, or different threshold values.\\ The computational complexity for the distribution of shares in the modified scheme and for reconstructing the secret is again in the same bounds, as the initial CTSS. Because as shown in Lemmas~\ref{lemma runtime sharing} and~\ref{lemma runtime reconstruction} sharing and reconstruction in the non-compartmented scheme remain in the same bounds. 
\subsection{Feasible Access Structures and Ideality}
\label{Our Contribution Feasible Access Structures}
Following the remarks of \cite{benaloh1990generalized} it is possible to construct the compartments for a CTSS to realize any access structure: Any formula in conjunctive normal form (CNF) can be mapped. Consider two shareholders $U_1$, $U_2$ in a $(2,2)$-secret sharing scheme. Neither $U_1$, nor $U_2$ are sufficient to retrieve the secret on their own. Therefore, the specific secret sharing scheme realizes the \textsc{and} operator. Whereas, two shareholders $U_1$, $U_2$ in a $(1,2)$-secret sharing scheme realize the \textsc{or} operator, because both can retrieve the secret on their own. Because both, the \textsc{and} and the \textsc{or} operator are possible to map, any formula in CNF can be mapped. The operators are not limited to be binary, but they can be $n$-ary. Consider the following scenario: There are four users $U_1$, $U_2$, $U_3$, and $U_4$, which can retrieve the secret if $U_1$ with either $U_2$ or $U_3$ and one of $U_2$, $U_4$ work together. The formula in CNF is $(U_1 \wedge (U_2 \vee U_3) \wedge (U_2 \vee U_4))$. The CTSS has three compartments $C_1 = \{U_1\}$, $C_2=\{U_2, U_3\}$, and $C_3=\{U_2, U_4 \}$, where $C_1$ holds only a single shareholder. The threshold is $t=3$ in the outer scheme, and $t=1$ in the inner schemes. In the given example the shareholder $U_2$ has to receive two independent and possibly unequal shares.\\
Previously, each level in the CTSS either could work as an \textsc{and} or an \textsc{or} operator. Now, by introducing the modifications to Shamir's scheme \textsc{and} and \textsc{or} operators can be used on the same level: Any crucial share $R$ is needed to retrieve a secret, therefore it is introducing another \textsc{and} operator. Two mutual redundant shares $S$ and $S'$ do not help in revealing the secret. Therefore, redundant shares introduce another kind of \textsc{or} operator. This \textsc{or} operator can be especially useful whenever a structure in the form $(A \wedge (B \vee C))$ appears, where $(B \wedge C)$ should not be allowed. The previous example can be mapped as a modified $(3,4)$-secret sharing scheme with $U_1$ receiving a crucial share, $U_2$ receiving a normal share, and both $U_3$ and $U_4$ receiving the same redundant share. The secret can be retrieved, if $U_1$, $U_2$, and one of $U_3$ or $U_4$ work together. In this scheme every shareholder gets one single share, therefore it is ideal. Furthermore, no CTSS is needed. This reduces both the computational complexity and the total amount of shares.\\
Using crucial and redundant shares is not useful in every case. For example, every secret sharing scheme, where only two shareholders participate can be constructed without crucial and redundant shares: A scheme consisting solely of two redundant shares can be realized by a $(1,2)$-secret sharing scheme. A scheme consisting solely of two crucial shares can be realized by a $(2,2)$-secret sharing scheme, or by the much simpler method displayed in Section~\ref{Introduction}. A scheme consisting of a crucial and a normal share can be mapped by a $(2,2)$-secret sharing scheme. In contrast, as soon as there are three shareholders the modifications can be useful. Consider a scheme consisting of a crucial share $R_1$, and two normal shares $S_1$ and $S_2$. The access structure is $(R_1 \wedge (S_1 \vee S2))$. This scheme, previously, could only be mapped by using a CTSS. Another example would be a scheme consisting of a normal share $S_1$ and two redundant shares $S_2$ and $S'_2$. This scheme previously needed a compartmented scheme like in the example before, because the access structure is $(S_1 \wedge (S_2 \vee S'_2))$.\\
The modifications cannot always help in constructing a scheme without compartments. Consider the access structure from~\cite{benaloh1990generalized}: ${((A \wedge B) \vee (B\wedge C) \vee (C\wedge D))}$. None of the shares can be a crucial share, because no share is present in all clauses. No pair of shares can be redundant: by making $A$ and $B$ redundant, the clause $(A\wedge B)=(A \wedge A)=(B \wedge B)$ would be unsatisfiable. By making $A$ and $C$ redundant, the clause $(A \wedge D)=(C \wedge D)$ would be implicitly introduced. By making $A$ and $D$ redundant, a new clause $(C\wedge A)=(C \wedge D)$ would be implicitly introduced. Therefore no set of shares containing $A$ can be redundant. Because of symmetry, the same problems occurs for the other combinations of shares. The given example, therefore, cannot be solved using the modified scheme without compartments. When using compartments, the resulting scheme is not ideal.
%
\section{Conclusions}
\label{Conclusions}
In this paper, we present some modifications to the secret sharing scheme of Shamir. The modifications allow defining redundant shares which give no additional advance in computing the secret if more than one mutual redundant shares are used. Further, crucial shares can be defined which are essential for retrieving the secret. We showed, that these modifications are easy to understand and implement, that the resulting schemes still are ideal, and that the computational complexity is not worse than in the original scheme. Further, the modifications are introduced to the CTSS, as proposed by Simmons. The modifications can help in constructing complex access structures, like for the scenarios described in the Introduction, by reducing the complexity of the access structure. Further, general access structures can be realized, where in some cases an optimal amount of distributed shares of one per shareholder can be achieved, i.e. the resulting scheme is ideal. In other cases, the computational complexity can be reduced compared to the naive approach.\\
Further modifications could be to introduce other operators. Especially the \textsc{xor} operator can be useful when trying to map access structures to a specific scheme and help in reducing the amount of distributed shares. But not all operators are always useful. Consider for example the \textsc{nor} operation, this would allow constructing a scheme consisting of multiple shares, where the secret can be calculated if no share is used. This is a contradiction to the initial idea behind secret sharing. Using compartmented secret sharing schemes clauses can be mapped where all shares or a specific number of shares are needed to reconstruct a secret. Introducing a random selector, where either the needed shares for calculating the secret are randomly chosen or depending on which shares are used to calculate the secret could be another modification to the scheme.

\bibliographystyle{IEEEtran}
\bibliography{secretsharing}

\end{document}